\documentclass{amsart}
\usepackage{amssymb,amsmath,amscd, amsthm,stmaryrd,verbatim, mathrsfs}

\newtheorem{Thm}{Main Theorem}
\newtheorem{theorem}{Theorem}[section]

\newtheorem{Prop}[theorem]{Proposition}

\theoremstyle{definition}
\newtheorem{Def}[theorem]{Definition}

\theoremstyle{remark}

\numberwithin{equation}{section}

\newcommand{\bb}{\mathbb}
\newcommand{\ms}{\mathscr}
\newcommand{\mr}{\mathrm}
\newcommand{\frk}{\mathfrak}

\usepackage{hyperref}
\begin{document}

\title{The $\mr O(1)$-Kepler Problems}
\author{Guowu Meng}

\address{Department of Mathematics, Hong Kong Univ. of Sci. and
Tech., Clear Water Bay, Kowloon, Hong Kong}

\email{mameng@ust.hk}

\subjclass[2000]{Primary 22E46, 22E70; Secondary 81S99, 51P05}

\date{May 7, 2008}


\keywords{Unitary Highest Weight Modules,  Kepler problems, Dual
Pairs, Theta Correspondences}

\begin{abstract} Let $n\ge 2$ be an integer.
To each irreducible representation $\sigma$ of $\mathrm O(1)$, an $\mathrm
{O}(1)$-Kepler problem in dimension $n$ is constructed and analyzed.
This system is super integrable and when $n=2$ it is equivalent to a
generalized MICZ-Kepler problem in dimension two. The dynamical
symmetry group of this system is $\widetilde {\mathrm{Sp}}_{2n}(\mathbb R)$
with the Hilbert space of bound states ${\mathscr H}(\sigma)$ being the
unitary highest weight representation of $\widetilde {\mathrm{Sp}}_{2n}(\mathbb R)$ with highest weight $$(\underbrace{-1/2, \cdots,
-1/2}_{n-1}, -(1/2+|\sigma|)),$$ which occurs at the right-most
nontrivial reduction point in the Enright-Howe-Wallach
classification diagram for the unitary highest weight modules. (Here
$|\sigma|=0$ or $1$ depending on whether $\sigma$ is trivial or
not.) Furthermore, it is shown that the correspondence
$\sigma\leftrightarrow \mathscr H(\sigma)$ is the theta-correspondence
for dual pair $(\mathrm{O}(1), \mathrm{Sp}_{2n}(\mathbb R))\subseteq \mathrm{Sp}_{2n}(\mathbb R)$.
\end{abstract}

\maketitle

\section {Introduction}
The Kepler problem is a well-known physics problem in dimension
three about two bodies which attract each other by a force
proportional to the inverse square of their distance. What is less
known about the Kepler problem is the fact that it is {\em super
integrable}\footnote{A
physics model is called {\em super integrable} if the number of
independent symmetry generators is bigger than the number of degree
of freedom. For the Kepler problem, the degree of freedom is $3$ and
the number of independent symmetry generators is $5$.} at both the classical and the quantum level, and belongs
to a big family of super integrable models.

One interesting such family is the family of \emph{generalized
MICZ-Kepler problems}, i.e., the family of MICZ-Kepler problems
\cite{MC70, Z68} and their high dimensional analogues \cite{meng05}.
The detailed dynamical symmetry analysis of this family of super
integrable models has recently been carried out in Refs. \cite{MZ07,
M07}. It is fair to say that the study of this family of super
integrable models has enriched both the field of super integrable
systems and the theory of unitary highest weight modules for real
non-compact Lie groups.

The purpose here is to construct and analyze yet another family of
super integrable models of the Kepler type. Recall that, in the
construction of generalized MICZ-Kepler problems in dimension $D$,
the canonical bundle
$$
\mr{Spin}(D-1)\to \mr{Spin}(D)\to \mr{S}^{D-1}
$$
plays a pivotal role. Here the corresponding bundle is
$$
\mr{O}(1)\to \mr{S}^{n-1}\to\bb R\mr{P}^{n-1}. $$ Later we shall
demonstrate that, in dimension two, a model constructed in this
paper is equivalent to a generalized MICZ-Kepler problem. That is
why the models constructed here are called \emph{$\mr O(1)$-Kepler
problems}.

Before stating our main result, let us fix some notations:
\begin{itemize}
\item $n$ --- an integer which is at least 2;
\item $\sigma$ --- an irreducible representation of $\mr{O}(1)$;
\item $|\sigma|$ --- an integer equal to $0$ (resp. $1$) if $\sigma$ is trivial (resp. nontrivial);
\item $\tilde {\mr U}(n)$ --- the nontrivial double cover of $\mr
U(n)$;
\item $\widetilde{\mr {Sp}}_{2n}(\bb R)$ --- the nontrivial double cover of
${\mr {Sp}}_{2n}(\bb R)$.
\end{itemize}
Note that there is a natural chain of groups:
$\mr{SO}(n)\subset \tilde{\mr U}(n)\subset \widetilde{\mr Sp}_{2n}(\bb R)$.

We are now ready to state the main  results on $\mr O(1)$-Kepler
problems.
\begin{Thm}\label{T:main} Let $n\ge 2$ be an integer, $\sigma$ an irreducible representation of $\mr{O}(1)$, and
$|\sigma|=0$ or $1$ depending on whether $\sigma$ is trivial or not.

For the $n$-dimensional $\mr O(1)$-Kepler problem with magnetic
charge $\sigma$, the following statements are true:

1) The bound state energy spectrum is
$$
E_I=-{1/2\over (I+{n\over 4}+{|\sigma|\over 2})^2}
$$ where $I=0$, $1$, $2$, \ldots

2) There is a natural unitary action of $\widetilde {\mr{Sp}}_{2n}(\bb R)$ on the Hilbert space ${\ms H}(\sigma)$ of negative-energy
states, which extends the manifest unitary action of $\mr{SO}(n)$.
In fact, ${\ms H}(\sigma)$ is the unitary highest weight module of
$\widetilde {\mr{Sp}}_{2n}(\bb R)$ with highest weight
$\left(-{1\over 2}, \cdots, -{1\over 2}, -({1\over
2}+|\sigma|)\right)$.

3) When restricted to the maximal compact subgroup $\tilde{\mr
U}(n)$, the above action yields the following orthogonal
decomposition of $\ms H(\sigma)$:
$$
{\ms H}(\sigma)=\hat\bigoplus _{I=0}^\infty\,{\ms H}_I(\sigma)
$$ where ${\ms H}_I(\sigma)$ is a model for the irreducible $\tilde {\mr U}(n)$-representation
with highest weight $\left(-{1\over 2}, \cdots, -{1\over 2},
-({1\over 2}+|\sigma|+2I)\right)$.

4) ${\ms H}_I(\sigma)$ in part 3) is the energy eigenspace with
eigenvalue $E_I$ in part 1).

5) The correspondence between $\sigma$ and $\ms H(\sigma)$ is just
the theta-correspondence\footnote{See Ref. \cite{Howe} for
details on reductive dual pairs and theta-correspondence.} for dual
pair $(\mr{O}(1), \mr{Sp}_{2n}(\bb R))\subseteq \mr{Sp}_{2n}(\bb R)$.
\end{Thm}
For readers who are familiar with the Enright-Howe-Wallach
classification diagram \cite{EHW82} for the unitary highest weight
modules, we would like to point out that, the unitary highest weight
module identified in part 2) of this theorem occurs at the rightmost
nontrivial reduction point of the classification diagram, and  its
$K$-type formula is \emph{multiplicity free} in view of part 3) of
this theorem.

In section \ref{S:model}, we introduce the models and show that when $n=2$
they are equivalent to the generalized MICZ-Kepler problems in
dimension two. In section
\ref{S:analysis}, we give a detailed analysis of the models and
finish the proof of main theorem \ref{T:main}.

Depending on the interests of the readers, one may view this paper
and its sequels either as a journey to discover new super integrable
systems or as an effort to better understand the geometry of those
Wallach representations which occur at the rightmost nontrivial
reduction point of the classification diagram.

I appreciate the effort of the referee for his or her careful
reading of the original manuscript and especially the comments which
yield a more clear presentation of this paper.

\section{The models}\label{S:model}
Let $n\ge 2$ be an integer, $\bb R^n_*=\bb R^n\setminus\{0\}$, and
$\sigma$ an irreducible unitary representation of $\mr O(1)$.
Consider the principal bundle
$$
\mr O(1)\to \bb R^n_*\to \widetilde{\bb RP^n}
$$
where $\widetilde{\bb RP^n}$ is the quotient space of  $\bb R^n_*$
under the equivalence relation $x\sim -x$. In terms of the polar
coordinates $(\rho, \Theta)$, the Riemannian metric on
$\widetilde{\bb RP^n}$ is $d\rho^2+\rho^2\,d\Theta^2$ where
$d\Theta^2$ is the standard round metric on $\bb RP^{n-1}$ .

Let $\gamma_\sigma$ be the associated vector bundle attached to
representation $\sigma$. It is clear that $\gamma_\sigma$ is a flat hermitian line bundle over
$\widetilde{\bb RP^n}$.

\begin{Def}\label{Def:1st}  Let $n\ge 2$ be an integer and
$\sigma$ an irreducible representation of $\mr O(1)$. The $\mr
O(1)$-Kepler problem in dimension $n$ with magnetic charge $\sigma$
is the quantum mechanical system for which the wave functions are
smooth sections of $\gamma_\sigma$, and the hamiltonian is
$$H=-{1\over 8\rho}\Delta{1\over \rho}-{1\over \rho^2}$$ where
$\Delta$ is the Laplace operator twisted by $\gamma_\sigma$ and
$\rho([x])=|x|$.
\end{Def}

Observe that, in dimension two, $\bb R^2_*$ and $\widetilde{\bb
RP^2}$ are diffeomorphic. We use $(r, \phi)$ to denote the polar
coordinates on $\bb R^2_*$ and $(\rho, \theta)$ to denote the polar
coordinates on $\widetilde{\bb RP^2}$. Let $\pi$: $\widetilde{\bb
RP^2}\to \bb R^2_*$ be the diffeomorphism such that $\pi(\rho,
\theta)=(\rho^2, 2\theta)$, then
$$
\pi^*(dr^2+r^2\,d\phi^2)=4\rho^2 (d\rho^2+\rho^2\,d\theta^2)\quad
\mbox{and}\quad \pi^*(\mr{vol}_{{\bb
R^2}_*})=4\rho^2\mr{vol}_{\widetilde{\bb RP^2}}.
$$

Let $\mu=0$ or $1/2$, and $\sigma_\mu$: $\mr O(1)=\bb Z_2\to \bb C$
be the group homomorphism which maps the generator of $\mr O(1)$ to
$(-1)^{2\mu}$. Let $\gamma(\mu)$ be the pullback of
$\gamma_{\sigma_\mu}$ by $\pi^{-1}$. It is clear that $\gamma(\mu)$ is a flat hermitian line
bundle over $\bb R^2_*$.

Recall from the appendix of Ref. \cite{meng07c} that \emph{the
generalized MICZ-Kepler problem in dimension two with magnetic
charge $\mu$} is the quantum mechanical system for which the wave
functions are smooth sections of $\gamma(\mu)$, and the hamiltonian
is
$$\hat h=-{1\over 2}\Delta-{1\over r}$$ where $\Delta$ is
the Laplace operator twisted by $\gamma(\mu)$ and $r(x)=|x|$. We are now
ready to state the following
\begin{Prop} The generalized MICZ-Kepler problem in dimension
two with magnetic charge $\mu$ is equivalent to the $\mr
O(1)$-Kepler problem in dimension two with magnetic charge
$\sigma_\mu$.
\end{Prop}
\begin{proof}
Let $\Psi_i$ ($i=1$ or $2$) be a wave-section for the generalized
MICZ-Kepler problem in dimension two with magnetic charge $\mu$, and
$$\psi_i(\rho, \theta):=2\rho\pi^*(\Psi_i)(\rho,
\theta)=2\rho\Psi_i(\rho^2,2\theta).$$ Then it is not hard to see
that
$$\displaystyle \int_{\widetilde{\bb
RP^2}}\overline{\psi_1}\psi_2\mr{vol}_{\widetilde{\bb
RP^2}}=\displaystyle \int_{\widetilde{\bb
RP^2}}\overline{\pi^*(\Psi_1)}\pi^*(\Psi_2) \pi^*(\mr{vol}_{\bb
R^2_*})=\displaystyle \int_{\bb R^2_*}\overline{\Psi_1}
\Psi_2\mr{vol}_{\bb R^2_*}$$ and
\begin{eqnarray}\displaystyle
\displaystyle\int_{\widetilde{\bb RP^2}}\overline{\psi_1}H\psi_2
\mr{vol}_{\widetilde{\bb RP^2}} &=&
\displaystyle\int_{\widetilde{\bb
RP^2}}\overline{\pi^*(\Psi_1)}\,{1\over \rho}H\rho\,\pi^*(\Psi_2)
\,\pi^*(\mr{vol}_{\bb R^2_*})\cr&=& \displaystyle\int_{\bb
R^2_*}\overline{\Psi_1}\,\hat h\,\Psi_2 \,\mr{vol}_{\bb R^2_*}.
\nonumber
\end{eqnarray}Here we have used the fact that
\begin{eqnarray}
{1\over \rho}H\rho &=& -{1\over 8\rho^2}\left({1\over
\rho}\partial_\rho\rho\partial_\rho+{1\over
\rho^2}\partial_\theta^2\right)-{1\over \rho^2}\cr &=& -{1\over
2}\left({1\over r}\partial_rr\partial_r+{1\over
r^2}\partial_\phi^2\right)-{1\over r}=\hat h.\nonumber
\end{eqnarray}

\end{proof}
We end this section with an alternative definition for the
$\mr{O}(1)$-Kepler problems.

\begin{Def}[Alternative Definition]\label{Def:2nd}
Let $n$, $\sigma$ be as in definition \ref{Def:1st}, and
$|\sigma|=0$ or $1$ depending on whether $\sigma$ is trivial or not.
The $\mr O(1)$-Kepler problem in dimension $n$ with magnetic charge
$\sigma$ is the quantum mechanical system for which the wave
functions are smooth complex-valued functions $\psi$ on $\bb R^n_*$
satisfying condition $\psi(-x)=(-1)^{|\sigma|} \psi(x)$, and the
hamiltonian is
\begin{eqnarray}\label{H:2nd}
H=-{1\over 8r}\Delta{1\over r}-{1\over r^2}\end{eqnarray} where
$\Delta$ is the Laplace operator on $\bb R^n_*$ and $r(x)=|x|$.
\end{Def}

\section{The dynamical symmetry analysis}\label{S:analysis}
We use definition \ref{Def:2nd}. Let $\psi$ be the eigenfunction of
$H$ in Eq. (\ref{H:2nd}) with eigenvalue $E$, so $\psi$ is square
integrable with respect to the Lebesque measure $d\mu$,
$\psi(-x)=(-1)^{|\sigma|} \psi(x)$, and
\begin{eqnarray}\label{E:eigenvalue}
\left(-{1\over 8r}\Delta{1\over r}-{1\over r^2}\right)\psi=E\psi.
\end{eqnarray}
We shall solve this eigenvalue problem by separating the angles from
the radius. The branching rule for $({\mr {SO}}(n), {\mr {SO}}(n-1))$
plus the Fubini Reciprocity law together imply that, as modules of
$\mr {SO}(n)$,
$$
L^2({\mr S}^{n-1})=\hat\bigoplus_{l=0}^\infty {\ms R}_l
$$ where ${\ms R}_l$ is the irreducible and unitary representation of $\mr{SO}(n)$ with the
highest weight $(l, 0, \cdots, 0)$.

Let $\{Y_{l{\bf m}}\mid {\bf m}\in {\mathcal I}(l)\}$ be a minimal
spanning set for ${\ms R}_l$. Write $\psi(x)=\tilde R_{kl}(r)Y_{l\bf
m}(\Omega)$ where $Y_{l\bf m}(\Omega)\in {\ms R}_l$. Note that
condition $\psi(-x)=(-1)^{|\sigma|} \psi(x)$ is equivalent to
equation $l\equiv |\sigma| \mod 2$. After separating out the angular
variables, Eq. (\ref{E:eigenvalue}) becomes
\begin{eqnarray}
\left({1\over 8}\left( -{1\over
r^n}\partial_rr^{n-1}\partial_r{1\over r}+{l^2+(n-2)l\over
r^4}\right)-{1\over r^2}\right)\tilde R_{kl}=E\tilde R_{kl}.
\end{eqnarray} where $\tilde R_{kl}\in L^2({\bb R}_+, r^{n-1}\,dr)$.
Let $R_{k{l\over 2}}(t)=\tilde R_{kl}(\sqrt t)/\sqrt t$, then we
have $ R_{k{l\over 2}}\in L^2({\bb R}_+, t^{n\over 2}\,dt)$ and
\begin{eqnarray}
\left( -{1\over 2t^{n\over 2}}\partial_t t^{n\over
2}\partial_t+{({l\over 2})^2+({n\over 2}-1){l\over 2}\over
2t^2}-{1\over t}\right) R_{k{l\over 2}}=E R_{k{l\over 2}}.
\end{eqnarray}
By quoting results from appendix A, we have
\begin{eqnarray}
E_{kl}=-{1/2\over (k+{l\over 2}+{n\over 4}-1)^2}
\end{eqnarray} where $k=1, 2, 3, \cdots$.
Let $I=k-1+{l-|\sigma|\over 2}$, then the {\em bound energy spectrum} is
\begin{eqnarray}
E_I=-{1/2\over (I+{n\over 4}+{|\sigma|\over 2})^2} \end{eqnarray}
where $I=0, 1, 2, \cdots$; since $\tilde R_{kl}(r)=rR_{k{l\over
2}}(r^2)$, we have
$$
\tilde R_{kl}(r)=c(k,l/2)r^{l+1}L^{l+{n\over 2}-1}_{k-1}\left({2\over
I+{n\over 4}+{|\sigma|\over 2}}r^2\right)\exp\left(-{r^2\over
I+{n\over 4}+{|\sigma|\over 2}}\right).
$$
This proves part 1) of the main theorem.

For each integer $I\ge 0$, we let ${\ms H}_I(\sigma)$ be the linear
span of
$$
\{\tilde R_{kl}Y_{l{\bf m}}\mid l\equiv |\sigma| \mod 2,\,{\bf
m}\in{\mathcal I}(l),\, k-1+{l-|\sigma|\over 2}=I\},
$$ then
$$
{\ms H}_I(\sigma)\cong \bigoplus_{k=0}^{I} {\ms R}_{2k+|\sigma|}
$$ is the eigenspace of $H$ with eigenvalue $E_I$, and the Hilbert
space of bound states admits the following orthogonal decomposition
into the eigenspace of $H$:
$$
{\ms H}(\sigma)=\hat\bigoplus_{I=0}^\infty {\ms H}_I(\sigma).
$$
Part 4) of the main theorem is then clear. We shall show that ${\ms
H}_I(\sigma)$ is the irreducible representation of $\tilde{\mr
U}(n)$ with highest weight $(-{1\over 2},\cdots,-{1\over 2},
-{1\over 2}+|\sigma|+2I)$ and ${\ms H}(\sigma)$ is the unitary
highest weight representation of $\widetilde{\mr{Sp}}_{2n}(\bb R)$
with highest weight $(-{1\over 2},\cdots,-{1\over 2}, -{1\over
2}+|\sigma|)$. To do that, we need to twist the Hilbert space of
bound states and the eigenspaces.
\subsection{Twisting}
Let $n_I=I+{n\over 4}+{|\sigma|\over 2}$ for each integer $I\ge 0$.
For each $\psi_I\in {\ms H}_I$, as in Refs. \cite{Barut71,MZ07}, we
define its twist $\tilde \psi_I$ by the following formula:
\begin{eqnarray}
\tilde \psi_I(x)=c_I{1\over |x|}\psi_I(\sqrt{n_I\over 2}x)
\end{eqnarray} where $c_I>0$ is the unique constant such that
\begin{eqnarray}\label{isometry}\int|\tilde \psi_I|^2=\int|\psi_I|^2.\end{eqnarray} Since
$$
\left(-{1\over 8r}\Delta{1\over r}-{1\over
r^2}\right)\psi_I(x)=E_I\psi_I(x),
$$ after re-scaling: $x\to \sqrt{n_I\over 2}x$, we have
$$
\left(-{(2/n_I)^2\over 8r}\Delta{1\over r}-{2/n_I\over
r^2}\right)\psi_I(\sqrt{n_I\over 2}x)=E_I\psi_I(\sqrt{n_I\over 2}x),
$$ or
$$
\left(-{1\over
2}\Delta-{2n_I}\right)\tilde\psi_I(x)=n_I^2E_Ir^2\tilde\psi_I(x)=-{1\over
2} r^2\tilde\psi_I(x).
$$ Then
\begin{eqnarray}\label{osc}
\left(-{1\over 2}\Delta+{1\over 2} r^2\right)\tilde
\psi_I&=&2n_I\tilde \psi_I\cr &= &(2I+|\sigma|+{n\over 2})\tilde
\psi_I.
\end{eqnarray}
We use $\tilde {\ms H}_I(\sigma)$ to denote the span of all such
$\tilde\psi_I$'s, $\tilde {\ms H}(\sigma)$ to denote the Hilbert
space direct sum of $\tilde{\ms H_I}(\sigma)$. We write the linear
map sending $\psi_I$ to $\tilde\psi_I$ as $\tau$: ${\ms
H}(\sigma)\to \tilde{\ms H}(\sigma)$. In view of Eqs. (\ref{isometry}) and (\ref{osc}), it is easy to see that
$\tau$ is an linear isometry; and one can also check easily that
$\tilde{\ms H}_I(\sigma)$ is the $(2I+|\sigma|)$-th energy
eigenspace of the $n$-dimensional isotropic harmonic isolator with
hamiltonian $-{1\over 2}\Delta+{1\over 2} r^2$.

By quoting  results from appendix \ref{Oscillator} on the isotropic
harmonic oscillators, we know that $\tilde{\ms H}_I(\sigma)$ is a
model for the irreducible $\tilde{\mr U}(n)$-representation with
highest weight
$$\left(-{1\over 2}, \cdots, -{1\over 2}, -({1\over
2}+|\sigma|+2I)\right),$$ and $\tilde {\ms H}(\sigma)$ is the
unitary highest weight module of $\widetilde {\mr{Sp}}_{2n}(\bb R)$
with highest weight $\left(-{1\over 2}, \cdots, -{1\over 2},
-({1\over 2}+|\sigma|)\right)$. In view of that fact that $\tau$ is
an isometry, by pulling back the action of $\widetilde {\mr{Sp}}_{2n}(\bb R)$ on $\tilde {\ms H}(\sigma)$ via $\tau$, we get the action of
$\widetilde {\mr{Sp}}_{2n}(\bb R)$ on ${\ms H}(\sigma)$. Then we have
parts 2), 3) and 5) of the main theorem proved.

\appendix
\section{Radial Schr\"{o}dinger Equation}
Let $l\ge 0$, $m>0$ be half integers, and $l'=l+{m\over 2}-1$. We
are interested in finding a nonzero $R_{kl}\in L^2({\bb R}_+,
r^m\,dr)$ satisfying the radial Schr\"{o}dinger equation:
\begin{eqnarray}\label{rSchEq}
\left(-{1\over 2r^m}\partial_r r^m\partial_r+{l^2+(m-1)l\over
2r^2}-{1\over r}\right)R_{kl}(r)=E_{k,l}R_{kl}(r).
\end{eqnarray} for some real number $E_{k,l}$.

Solving this differential equation by the {\em power series method},
we can see that, for $R_{kl}(r)$ to be square integrable with
respect to measure $r^m\,dr$, we must have
$$
E_{k,l}=-{1/2\over (k+l')^2}
$$ for some positive integer $k$. With this value for $E_{kl}$ in mind, we
plug
$$R_{kl}(r)=r^{-{m\over 2}}y_{kl'}(r)\exp\left(-{r\over k+l'}\right)$$
into Eq. (\ref{rSchEq}), and get
\begin{eqnarray}\label{E:lugree}
\left( {d^2\over dr^2} -{2\over k+l'} {d\over dr}+\left[{2\over
r}-{l'(l'+1)\over r^2}\right]\right)y_{kl'}(r)=0.\end{eqnarray} Here
$y_{kl'}(r)$ is square integrable with respect to measure
$\exp\left(-{2r\over k+l'}\right)\,dr$.

Recall that, for nonnegative integers $n$ and $k$, generalized
Laguerre polynomial $L^k_n$ is defined to be
$$L^k_n(x)={e^x x^{-k}\over n!}
{d^n\over dx^n}\left(e^{-x}x^{n+k}\right).$$

Solving Eq. (\ref{E:lugree}) in terms of power series under the
square integrability condition, we arrive at the general solution of
the following form:
$$
y_{kl'}(r)=c(k,l) r^{l'+1}L^{2l'+1}_{k-1}\left({2\over
k+l'}r\right).
$$ Here $c(k,l)$ is a constant, which
can be uniquely determined by requiring $c(k,l)>0$ and
$\int_0^\infty |R_{kl}(r)|^2 r^{m}dr=\int_0^\infty |y_{kl'}(r)|^2
\exp\left(-{2r\over k+l'}\right)\,dr=1$.

\section{Isotropic Harmonic Oscillators}\label{Oscillator}
The purpose here is to spell out the details on the quantum
isotropic harmonic oscillator in dimension $n$. We work in the
Schr\"{o}dinger picture, then the hamiltonian is
$$
H={1\over 2}\left(-\Delta+r^2\right).
$$ Here $r=\sqrt{\sum _{i=1}^n(x^i)^2}$ and
$\Delta=\sum_{i=1}^n\partial_i^2$.

Let $a_k={1\over \sqrt 2}(x_k+\partial_k)$ and $a_k^\dag$ be the
hermitian adjoint of $a_k$. $a_k$'s are called annihilation
operators and $a_k^\dag$'s are called the creation operators. It is
a standard fact that all bound states are created from the ground
state $|\Omega\rangle$ via the creation operators. By definition, a
$k$-th {\em excited state} is a state created from the ground state
via a degree $k$ polynomial in creation operators. For example,
$(a_1^\dag-3a_2^\dag)|\Omega\rangle$ is a first excited state, and
$\left(a_1^\dag a_2^\dag+3(a_2^\dag)^2\right)|\Omega\rangle$ is a
second excited state.

Introducing operators
\begin{eqnarray} -H_i =a_i^\dag a_i+{1\over 2} & & \hbox{for $1\le i\le
n$},\cr E_{-e^j+e^k} = a_j^\dag a_k & & \mbox{for $1\le j< k \le
n$},\cr E_{-e^j-e^k} = a_j^\dag a_k^\dag & & \mbox{for $1\le j< k\le
n$},\cr E_{-2e^j} = {1\over \sqrt 2}a_j^\dag a_j^\dag & & \mbox{for
$1\le j\le n$}.\nonumber
\end{eqnarray}It can be verified that the Hamiltonian can be written as
$$
H=-\sum_i H_i=\sum_{k=1}^na_k^\dag a_k+{n\over 2}\;;
$$ moreover, operators $H_i$, $E_{-e^j\pm e^k}$, $E_{-e^j\pm e^k}^\dag$
satisfy the commutation relations of a Cartan-Chevalley basis
for $\frk{sp}_{2n}(\bb R)$.

It is then clear that the ground state is non-degenerate and has
highest weight $\left(-{1\over 2}, \cdots, -{1\over 2}\right)$, the
1st excited states are degenerate and form the unitary highest
weight representation of $\frk{u}(n)$ with highest weight
$\left(-{1\over 2}, \cdots, -{1\over 2}, -{3\over 2}\right)$. In
general, the $I$-th excited states form the unitary highest weight
representation of $\frk{u}(n)$ with highest weight $\left(-{1\over
2}, \cdots, -{1\over 2}, -({1\over 2}+I)\right)$.

Under $\frk{sp}_{2n}(\bb R)$, the Hilbert space of bound states
splits into two irreducible components: the one consisting of states
with even number of particles and the one consisting of states with
odd number of particles.


\begin{thebibliography}{99}

\bibitem{MC70}
H. McIntosh and A. Cisneros, Degeneracy in the presence of a
magnetic monopole, {\em J. Math. Phys.} {\bf 11} (1970), 896-916.

\bibitem{Z68}
D. Zwanziger, Exactly soluble nonrelativistic model of particles
with both electric and magnetic charges, {\em Phys. Rev.} {\bf 176}
(1968), 1480-1488.


\bibitem{meng05} G. W. Meng, MICZ-Kepler
problems in all dimensions. {\em J. Math. Phys.} {\bf 48} (2007),
032105. {\em E-print}, arXiv:math-ph/0507028.


\bibitem{MZ07}
G. W. Meng and R. B. Zhang, Generalized MICZ-Kepler Problems and
Unitary Highest Weight Modules. {\em E-print},
arXiv:math-ph/0702086.

\bibitem{M07}
G. W. Meng, Generalized MICZ-Kepler Problems and Unitary Highest
Weight Modules -- II. {\em E-print}, arXiv:0704.2936.

\bibitem{EHW82} T. Enright, R. Howe and N. Wallach, A
classification of unitary highest weight modules, {\em
Representation theory of reductive groups}, Progress in Math. {\bf
40}, Birkh\"{a}user (1983), 97-143.

\bibitem{Howe}
R. Howe, Dual pairs in physics: harmonic oscillators, photons,
electrons, and singletons. {\em Lectures in Appl. Math.} {\bf 21}
(1985), Amer. Math. Soc., Providence, RI.



\bibitem{meng07c}
G. W. Meng, The Representation Aspect of the Generalized Hydrogen
Atoms. To appear in {\em J. of Lie Theory}. {\em E-print},
arXiv:0704.3107.


\bibitem{Barut71} A. Barut and G. Bornzin,
$\mr{SO}(4,2)$-Formulation of the Symmetry Breaking in Relativistic
Kepler Problems with or without Magnetic Charges, {\em J. Math.
Phys.} {\bf 12} (1971), 841-843.




\end{thebibliography}
\end{document}